\documentclass[twocolumn]{aastex6}
\usepackage{amsmath,amstext}
\usepackage[T1]{fontenc}
\usepackage{apjfonts} 
\usepackage[figure,figure*]{hypcap}
\usepackage[tight]{units}
\usepackage{natbib}

\newcolumntype{K}[1]{>{\centering\arraybackslash}p{#1}}

\shorttitle{UHECRs and extragalactic gamma-ray sources}
\shortauthors{The Pierre Auger Collaboration}

\begin{document}

\title{Indication of anisotropy in arrival directions of ultra-high-energy cosmic rays through comparison to the flux pattern of extragalactic gamma-ray sources}

\AuthorCallLimit=-1
\author{The~Pierre~Auger~Collaboration}
\affil{see the end matter for the full list of authors.\vspace{0.5cm}}
\email{auger_spokespersons@fnal.gov}

\published{in ApJL as DOI:10.3847/2041-8213/aaa66d}

\begin{abstract}
A new analysis of the dataset from the Pierre Auger Observatory provides evidence for anisotropy in the arrival directions of ultra-high-energy cosmic rays on an intermediate angular scale, which is indicative of excess arrivals from strong, nearby sources. The data consist of 5514 events above $\unit[20]{EeV}$ with zenith angles up to $\unit[80]{^\circ}$ recorded before 2017 April 30. Sky models have been created for two distinct populations of extragalactic gamma-ray emitters: active galactic nuclei from the second catalog of hard \textit{Fermi}-LAT sources (2FHL) and starburst galaxies from a sample that was examined with \textit{Fermi}-LAT. Flux-limited samples, which include all types of galaxies from the \textit{Swift}-BAT and 2MASS surveys, have been investigated for comparison. The sky model of cosmic-ray density constructed using each catalog has two free parameters, the fraction of events correlating with astrophysical objects and an angular scale characterizing the clustering of cosmic rays around extragalactic sources. A maximum-likelihood ratio test is used to evaluate the best values of these parameters and to quantify the strength of each model by contrast with isotropy. It is found that the starburst model fits the data better than the hypothesis of isotropy with a statistical significance of $\unit[4.0]{\sigma}$, the highest value of the test statistic being for energies above $\unit[39]{EeV}$. The three alternative models are favored against isotropy with $\unit[2.7-3.2]{\sigma}$ significance. The origin of the indicated deviation from isotropy is examined and prospects for more sensitive future studies are discussed.  
\end{abstract}


\section{Search for UHECR anisotropies}

Identifying the sources of ultra-high-energy cosmic rays (UHECRs) has been a prime goal of particle astrophysics for decades.  The challenge is great, because the flux falls rapidly with increasing energy, and because UHECRs have a mixed mass composition \citep{PhysRevD.90.122006, 2016PhLB..762..288A} so that some or all of them experience substantial magnetic deflections. 
Many scenarios have been proposed involving different populations of host galaxies.  
In this Letter, we investigate whether intermediate-scale\footnote{``Intermediate'' denotes hereafter angular scales larger than the experimental resolution, ${\sim}\unit[1]{^\circ}$, and smaller than large-scale patterns, ${\gtrsim}\unit[45]{^\circ}$.} anisotropies in UHECR arrival directions are associated with either or both of two prominent classes of extragalactic sources detected by \textit{Fermi}-LAT -- active galactic nuclei (AGNs) and starburst galaxies (SBGs) -- using the gamma-ray luminosity or its surrogate (radio emission for SBGs) as a proxy for the relative luminosity of each source in UHECRs.

The rate of energy production of UHECRs determined from observations above $\unit[10^{18}]{eV}$ is close to $\unit[10^{45}]{erg\,Mpc^{-3}\,yr^{-1}}$ \citep{Unger:2015laa}. Based on the {\it Fermi}-LAT survey, \cite{2010ApJ...724.1366D} argue that AGNs and SBGs match such rates in the gamma-ray band.
Due to the low density of detected SBGs and AGNs, and the attenuation of UHECR flux with increasing distance \citep[GZK effect,][]{1966PhRvL..16..748G,1966JETPL...4...78Z}, a few objects would be expected to dominate the local flux, naturally producing an intermediate-scale anisotropy if these sources contribute a sufficient fraction of the UHECR flux. 

The AGN and SBG populations are well-motivated physically.  AGNs are favored source candidates because their jets and radio lobes satisfy the Hillas criterion for shock acceleration \citep{1984ARA&A..22..425H}. SBGs -- being loci of intense star formation -- potentially have increased rates of extreme events associated with the deaths of short-lived, massive stars, such as gamma-ray bursts, hypernovae, and magnetars  \cite[see e.g.][]{2016arXiv161000944B, 2016SSRv..202..111P}. Their winds have also been proposed as possible reacceleration sites \citep{1999PhRvD..60j3001A}.

The analysis presented here is an advance in several ways. First, \textit{Fermi}-LAT observations of gamma rays from two extragalactic populations provide us with possible ansatzes for the relative UHECR fluxes from source candidates. That information makes the present analyses potentially more sensitive than previous studies based solely on the source direction. Second, thanks to our improved knowledge of the energy-dependent composition, we can now account more accurately for the relative attenuation of fluxes from distant sources. Third, thanks to the significant increase in exposure of the Pierre Auger Observatory with respect to previous analyses, the data can reveal more subtle patterns. 

\section{UHECR dataset}
UHECRs are detected at the Pierre Auger Observatory ~\citep[Argentina, latitude $35.2\,^\circ$~S, longitude $69.5\,^\circ$~W;][]{ThePierreAuger:2015rma} through the extensive air showers they induce in the atmosphere. Air showers are detected on the ground with an array of 1,600 water-Cherenkov detectors with a duty cycle of nearly $\unit[100]{\%}$. Twenty-four fluorescence telescopes map, during dark nights (duty cycle of ${\sim}\unit[15]{\%}$), the longitudinal profile of each shower via the nitrogen fluorescence produced dominantly by the electromagnetic cascade. The combination of both techniques provides the array with an energy scale insensitive to primary mass assumptions and air-shower simulation uncertainties. The systematic uncertainty in the energy scale is estimated to be $\unit[14]{\%}$~\citep{energyscale}.

Events above $\unit[20]{EeV}$ recorded between 2004 January 1 and 2017 April 30 are used in this analysis. Above $\unit[20]{EeV}$, both `vertical showers' \citep[zenith angle $\theta<60\,^\circ$,][]{2010NIMPA.613...29A} and `inclined showers' \citep[$60\,^\circ\leq\theta\leq80\,^\circ$,][]{2014JCAP...08..019P} trigger the array of detectors with $\unit[100]{\%}$ efficiency, the average angular resolution being below $\unit[1]{^\circ}$ and the statistical energy resolution being better than $\unit[12]{\%}$. 

Combining the vertical and inclined datasets, including unfolding correction factors as in \cite{2015JCAP...08..049P}, enables sky coverage over the declination range $\unit[-90]{^\circ}<\delta<+\unit[45]{^\circ}$. Using the same selection criteria as in \cite{2015ApJ...804...15A}, the total exposure for the period considered here is $\unit[89,720]{km^2\,sr\,yr}$.

\section{Source selection \& UHECR sky models}

\subsection{Extragalactic gamma-ray populations}

We extract our list of gamma-ray AGNs ($\gamma$AGNs hereafter) from the 2FHL catalog \citep{2016ApJS..222....5A}, which includes 360 sources detected by \textit{Fermi}-LAT above $\unit[50]{GeV}$. We study radio-loud objects within a $\unit[250]{Mpc}$ radius, yielding 17 blazars and radio galaxies. Their $\unit[50]{GeV}-\unit[2]{TeV}$ integral flux is used as a proxy for the UHECR flux. Given the distance of these objects, the gamma-ray absorption by the extragalactic background light \citep[e.g.][]{D11} is small.

The detections of seven SBGs have been reported using \textit{Fermi}-LAT data: NGC~253, M~82, NGC~4945, NGC~1068 \citep{2012ApJ...755..164A}, NGC~2146 \citep{2014ApJ...794...26T}, Arp~220 \citep{2016ApJ...821L..20P}, and Circinus \citep{2013ApJ...779..131H}. Their gamma-ray luminosity has been shown to scale almost linearly with their continuum radio flux \citep{2012ApJ...755..164A}. We thus adopt as a proxy for the UHECR flux of SBGs their continuum emission at $\unit[1.4]{GHz}$ for which a larger census exists. 

We select the 23 SBGs with a flux larger than $\unit[0.3]{Jy}$ among the 63 objects within $\unit[250]{Mpc}$ searched for gamma-ray emission by \citet{2012ApJ...755..164A}. Due to possible incompleteness of that list near the Galactic plane ($|b|<\unit[10]{^\circ}$) and in the southern sky ($\delta<\unit[-35]{^\circ}$), relevant SBGs could be missing from our selection. We checked however that our conclusions remain unchanged:\footnote{Test statistic for anisotropy within $\pm 1$ unit.} a) using all 63 objects listed in \citet{2012ApJ...755..164A}, b) using the catalog from \cite{2009arXiv0901.1775B} with 32 SBGs above $\unit[0.3]{Jy}$, c) adding the Circinus SBG absent from (a) and (b), d) using only the six SBGs reported in the 3FGL \citep[NGC~253, M~82, NGC~4945, NGC~1068, Circinus, NGC~2146;][]{2015ApJS..218...23A} and their $\unit[1-100]{GeV}$ integral flux as UHECR proxy.

\subsection{X-ray and infrared samples}

Following previous searches \citep{2015ApJ...804...15A}, we additionally study two flux-limited samples: \textit{Swift}-BAT sources up to $\unit[250]{Mpc}$, above a  flux of $\unit[13.4\times10^{-12}]{erg\,cm^{-2}\,s^{-1}}$, and sources from the 2MASS redshift survey \citep[2MRS catalog,][]{2012ApJS..199...26H} beyond $\unit[1]{Mpc}$, effectively taking out the Local Group as in \cite{2006MNRAS.368.1515E}. We use the $\unit[14-195]{keV}$ flux and K-band flux corrected for Galactic extinction as UHECR proxies for each of these surveys.

The X-ray sky observed by \textit{Swift}-BAT is dominated in flux by the nearby Centaurus~A, often considered as a prime UHECR source candidate \citep[e.g.][]{1996APh.....5..279R,2017arXiv170608229W}, with additional diffuse structures arising from both radio-loud and radio-quiet AGNs. This constitutes a different selection of AGNs from that performed for the $\gamma$-ray sample (radio-loud only), dominated by the radio galaxies Centaurus~A and M~87 within $\unit[20]{Mpc}$ and by the blazars Mrk~421 and Mrk~501 within $\unit[200]{Mpc}$.

The 2MRS infrared intensity traces the distribution of extragalactic matter, and includes both star-forming galaxies and AGNs. It is dominated by contributions from the nearby SBG NGC~253, close to the South Galactic pole, M~82, only observable from the Northern hemisphere, along with M~83 and NGC~4945, belonging to the same group of galaxies as Centaurus~A. Strong emission from Centaurus~A as well as cumulated emission from fainter objects, e.g. in the Virgo cluster, constitute distinctive features of the 2MRS sky model with respect to the SBG one.

\begin{figure*}[t]
\includegraphics[width=0.48\textwidth]{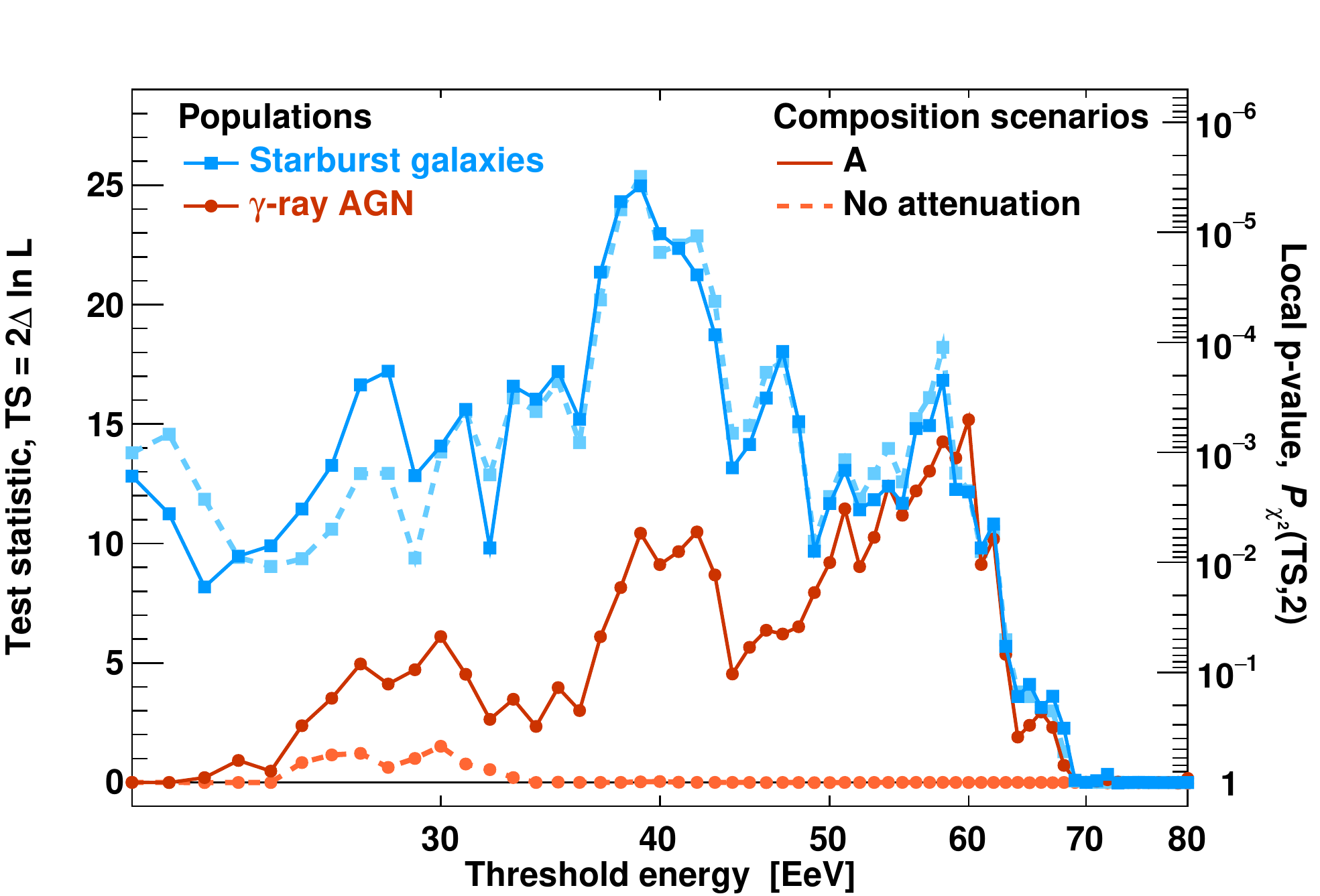}\hfill\includegraphics[width=0.48\textwidth]{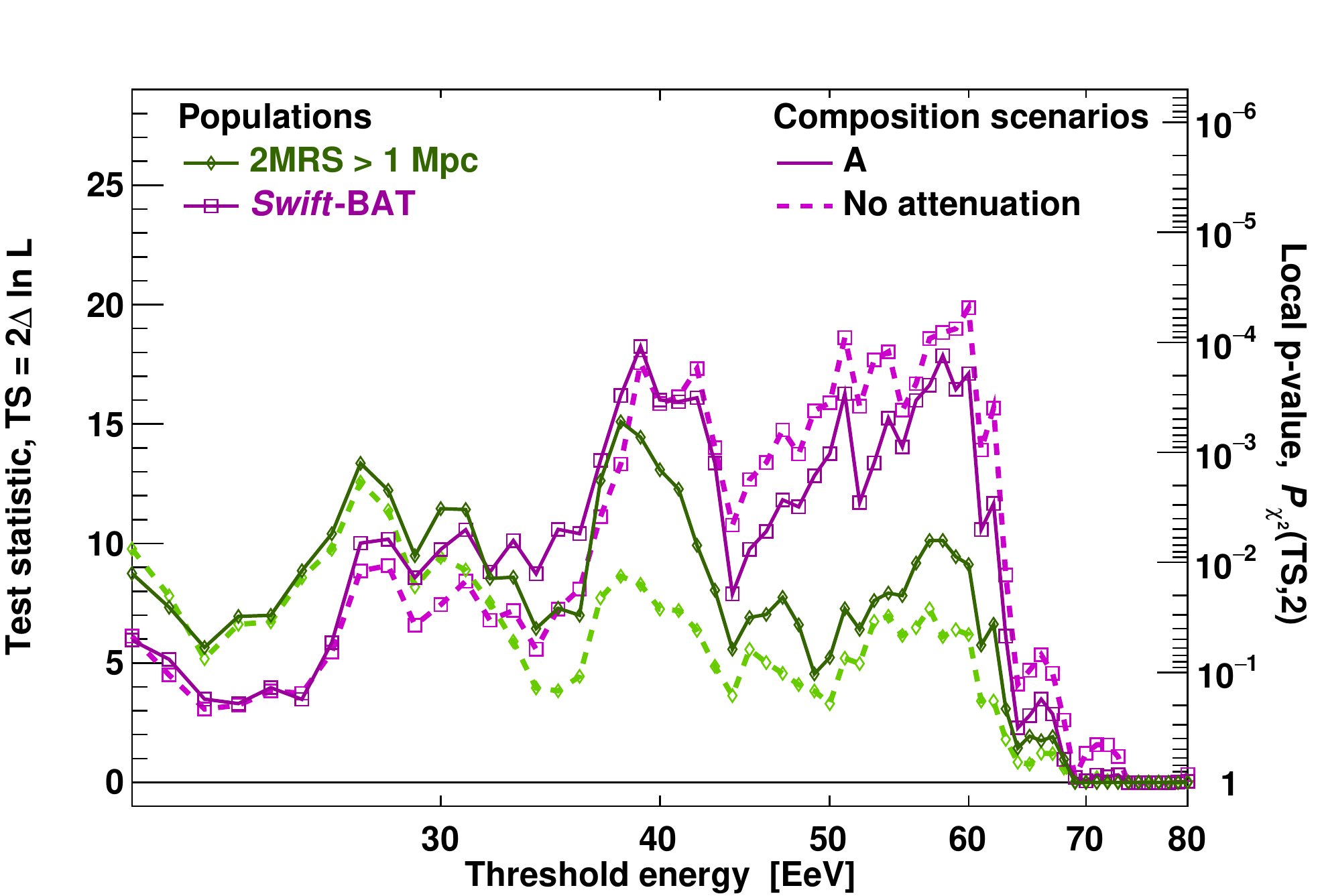}
\caption{TS scan over the threshold energy for SBGs and AGNs (\textit{Left}) and \textit{Swift}-BAT and 2MRS sources (\textit{Right}), including attenuation (light-dashed lines) or not (darker-solid lines).}
\label{Fig:Escan}
\end{figure*}

\subsection{Impact of attenuation}\label{Sec:compo}

We account for attenuation of UHECRs from distant objects (GZK effect) using a data-driven scenario that reproduces the composition and spectral constraints obtained by the Observatory \citep{2017JCAP...04..038A}. Assuming a homogeneous distribution of sources in the local Universe, it was shown that an interpretation of the air-shower data using the EPOS-LHC interaction model together with a hard injection index $\gamma=1$ at the sources (scenario A) best matches the data, accounting for propagation effects \citep{2012JCAP...10..007A,2016JCAP...05..038A}. We also consider two other scenarios matching the data reasonably well: EPOS-LHC with $\gamma=2$ (B) and Sibyll~2.1 with $\gamma=-1.5$ (C). These scenarios differ in the composition and maximum rigidities attainable at the sources. For each scenario and a chosen energy threshold, we evaluate the flux attenuation factor due to propagation for each source and correct its expected UHECR flux accordingly.

The two extragalactic gamma-ray populations under study and the relative weight of each source are provided in Table~\ref{tab:Pop}. The relative contributions  accounting for the directional exposure of the Observatory are shown in the last column. Because SBGs are mostly nearby, attenuation from them is much less important than from the more distant blazars in the $\gamma$AGN sample. Taking into account attenuation, ${\sim}\unit[90]{\%}$ of the accumulated flux from SBGs emerges from a ${\sim}\unit[10]{Mpc}$-radius region, while the radius goes up to ${\sim}\unit[150]{Mpc}$ for $\gamma$AGNs. For both the 2MRS and \textit{Swift}-BAT flux-limited samples, the $\unit[90]{\%}$ radius is ${\sim}\unit[70]{Mpc}$.

\section{Analysis and results}

\subsection{Maximum-likelihood analysis}

We build the UHECR sky model as the sum of an isotropic component plus the anisotropic contribution from the sources.  For the anisotropic component, each source is modeled as a Fisher distribution \citep[][]{1953RSPSA.217..295F}, the equivalent of a Gaussian on the sphere. Its distribution is centered on the coordinates of the source, the integral being set by its flux attenuated above the chosen energy threshold, and the angular width -- or search radius\footnote{Inverse square root of Fisher's concentration parameter.} -- being a free parameter common to all sources. No shift of the centroid position is considered, avoiding dependence on any particular model of the Galactic magnetic field in this exploratory study.   After mixing the anisotropic map with a variable fraction of isotropy, as in \cite{2010APh....34..314A}, the model map is multiplied by the directional exposure of the array and its integral is normalized to the number of events. The model map thus depends on two variables aimed at maximizing the degree of correlation with UHECR events: the fraction of all events due to the sources (anisotropic fraction) and the RMS angular separation between an event and its source (search radius) in the anisotropic fraction.

We perform an unbinned maximum-likelihood analysis, where the likelihood (L) is the product over the UHECR events of the model density in the UHECR direction. The  test statistic (TS) for deviation from isotropy is the likelihood ratio test between two nested hypotheses: the UHECR sky model and an isotropic model (null hypothesis). The TS is maximized as a function of two parameters: the search radius and the anisotropic fraction.  We repeat the analysis for a sequence of energy thresholds.

For a given energy threshold, we confirmed with simulations that the TS for isotropy follows a $\chi^2$ distribution with two degrees of freedom, as expected \citep{wilks1938}, directly accounting for the fit of two parameters of the model. As in \citet[][]{2015ApJ...804...15A}, we penalize the minimum p-value for a scan in threshold energy, by steps of $\unit[1]{EeV}$ up to $\unit[80]{EeV}$, estimating the penalty factor with Monte-Carlo simulations. The p-values are converted into significances assuming 1-sided Gaussian distributions.

\begin{figure*}[t]
\includegraphics[width=0.48\textwidth]{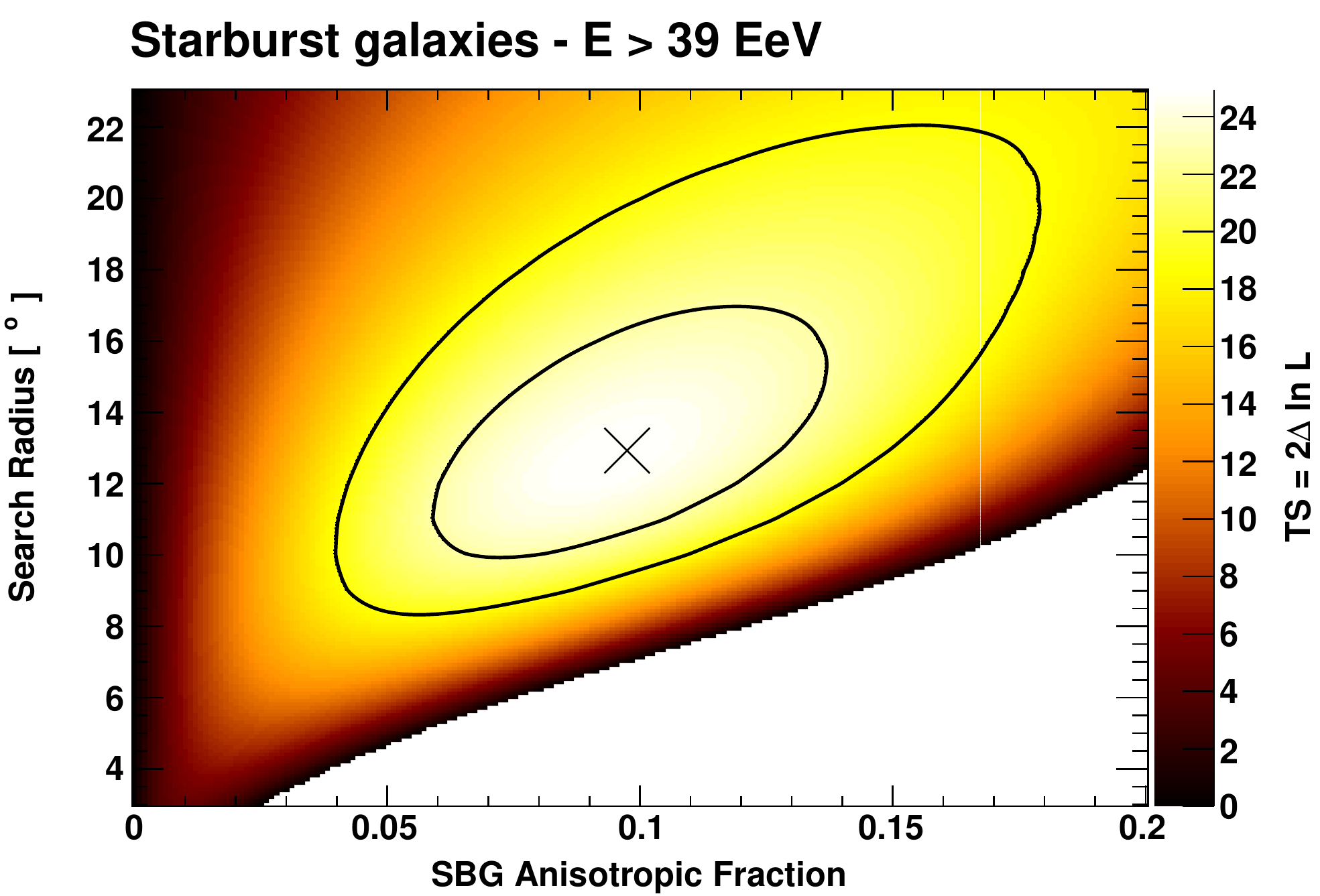}\hfill\includegraphics[width=0.48\textwidth]{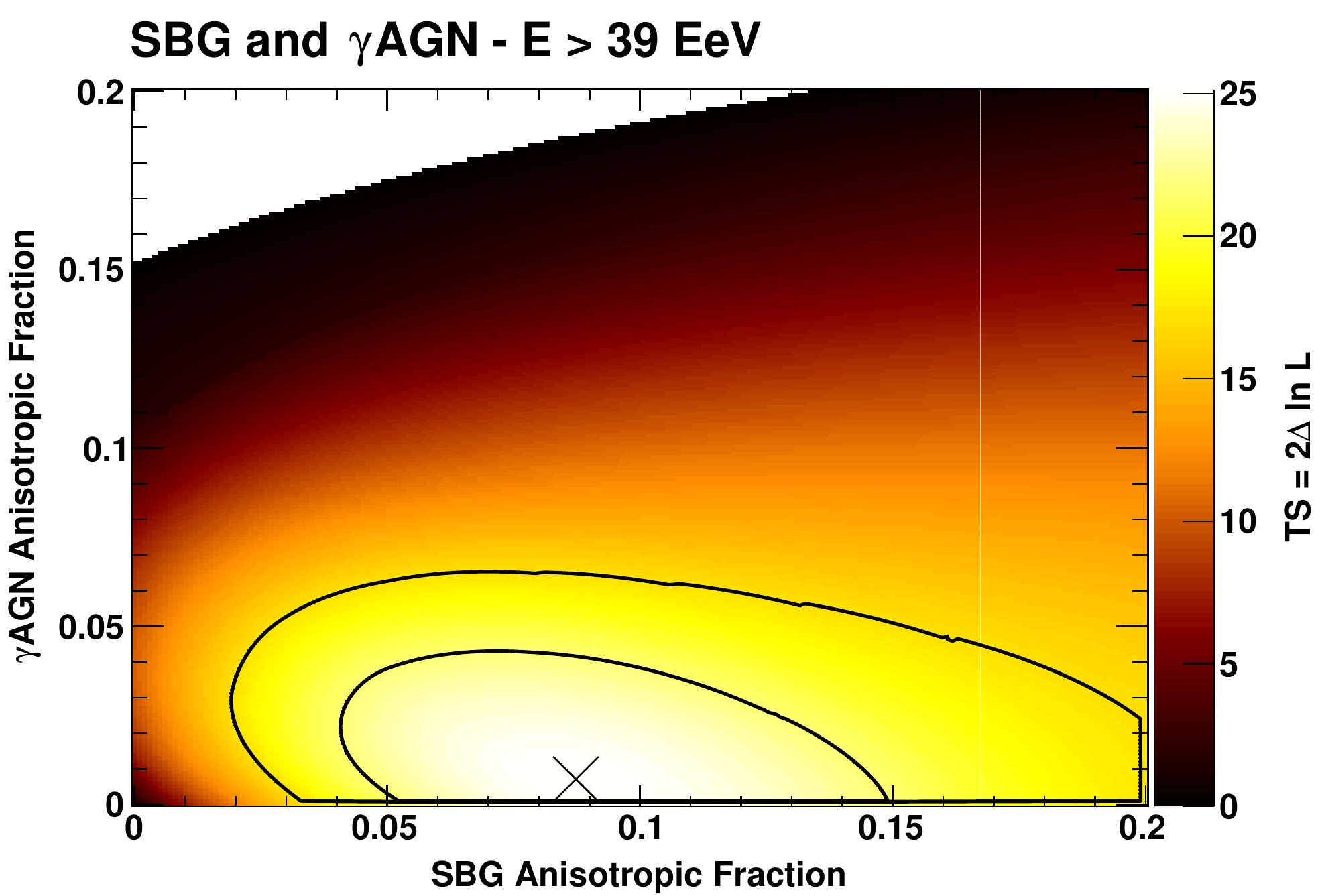}
\includegraphics[width=0.48\textwidth]{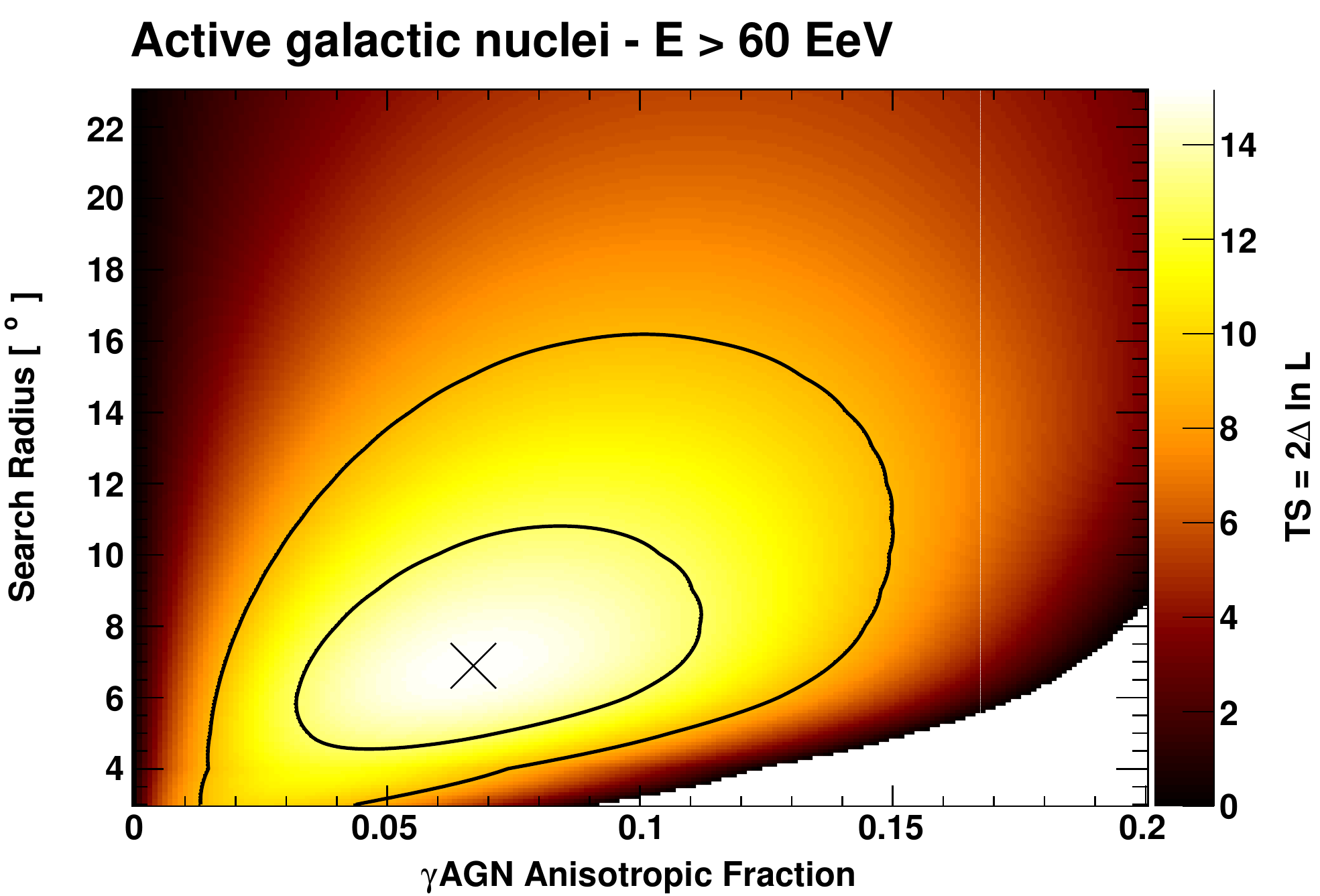}\hfill\includegraphics[width=0.48\textwidth]{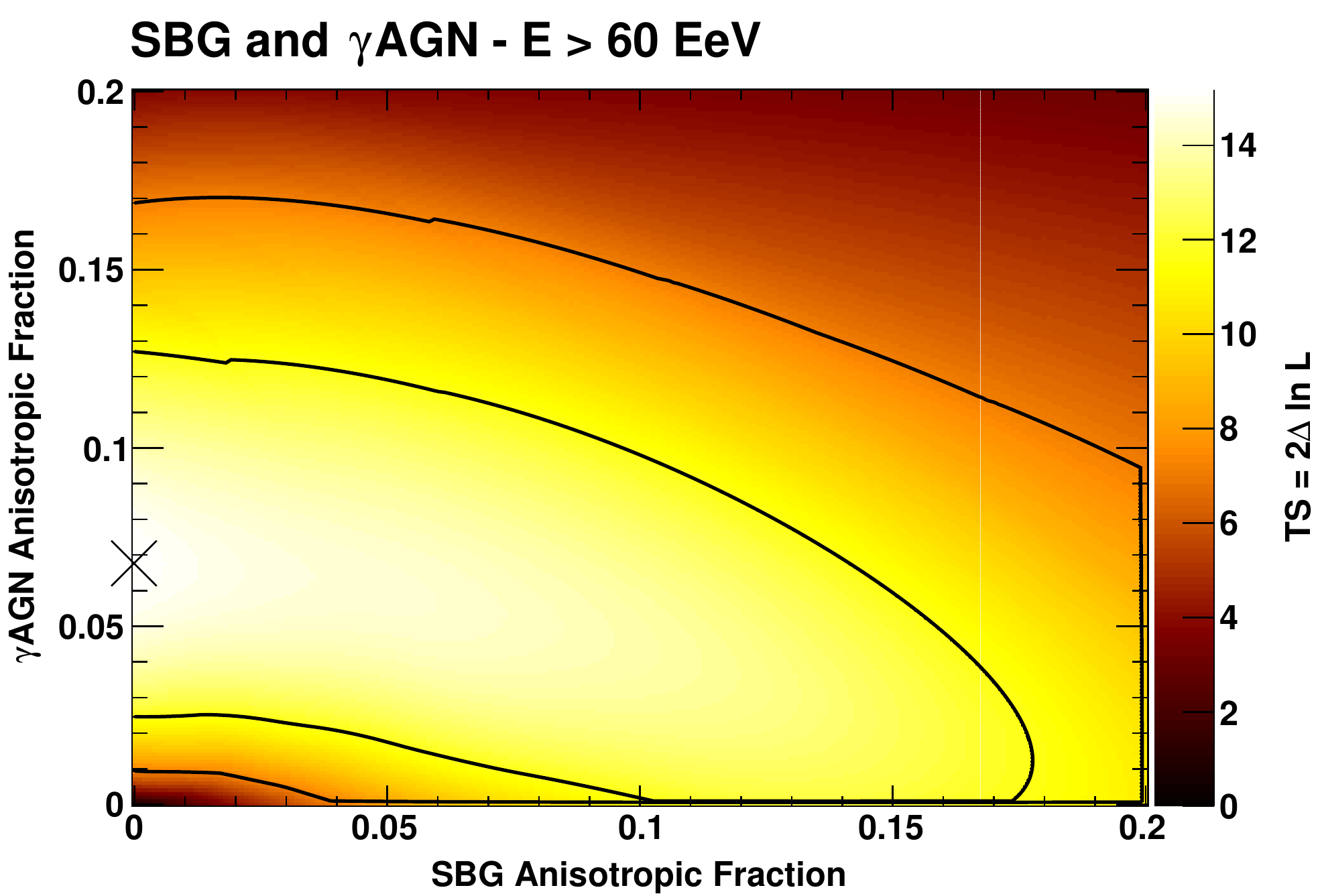}\\ 
\caption{TS profile above $\unit[39]{EeV}$ (\textit{Top}) and $\unit[60]{EeV}$ (\textit{Bottom}) over the fit parameters for SBG-only and $\gamma$AGN-only models (\textit{Left}) and for composite models including both SBGs and $\gamma$AGNs with the same free search radius (\textit{Right}). The lines indicate the $\unit[1-2]{\sigma}$ regions.}
\label{Fig:TSmaps}
\end{figure*}

\subsection{Single population against isotropy}

Previous anisotropy studies \citep[e.g.][]{2015ApJ...804...15A} have considered a scan in energy threshold starting at $\unit[40]{EeV}$, where the observed flux reaches half the value expected from lower-energy extrapolations, but as shown in Fig.~\ref{Fig:Escan}, there is a maximum in the significance close to this starting point. Therefore we have evaluated the TS down to $\unit[20]{EeV}$.

The TS is maximum for SBGs above $\unit[39]{EeV}$ (894 events), with or without attenuation. For $\gamma$AGNs, the TS is maximum above $\unit[60]{EeV}$ (177 events) after accounting for attenuation. As shown in Fig.~\ref{Fig:Escan}, left, attenuation  mildly impacts SBGs which are nearby: we obtain TS=24.9/25.5/25.7 for scenarios A/B/C, respectively. The impact is more pronounced for $\gamma$AGNs, a larger attenuation reducing contributions from distant blazars: we obtain a maximum TS of 15.2/9.4/11.9 for scenarios A/B/C. Shifting the energy scale within systematic uncertainties ($\pm\unit[14]{\%}$) affects the maximum TS by $\pm 1$ unit for $\gamma$AGNs, $\pm 0.3$ for SBGs.

Penalizing for the energy scan, the maximum TS obtained for SBGs and $\gamma$AGNs within scenario A correspond to $\unit[4.0]{\sigma}$ and $\unit[2.7]{\sigma}$ deviations from isotropy, respectively. As shown in Fig.~\ref{Fig:TSmaps}, left, the maximum deviation for $\gamma$AGNs is found at an angular scale of $\unit[7_{-2}^{+4}]{^\circ}$ and a $\unit[7 \pm 4]{\%}$ fraction of anisotropic events. For SBGs, a stronger deviation from isotropy is uncovered at an intermediate angular scale of $\unit[13_{-3}^{+4}]{^\circ}$ and an anisotropic fraction of $\unit[10 \pm 4]{\%}$. The systematic uncertainty induced by the energy scale and attenuation scenario is at the level of $\unit[0.3]{\%}$ for the anisotropic fraction and $\unit[0.5]{^\circ}$ for the search radius obtained with SBGs.

For \textit{Swift}-BAT and 2MRS sources attenuated within scenario A, we obtain maximum TS of 18.2 ($\unit[3.2]{\sigma}$) above $\unit[39]{EeV}$ and 15.1 ($\unit[2.7]{\sigma}$) above $\unit[38]{EeV}$, respectively (see Fig.~\ref{Fig:Escan}, right). These correspond to values of the best-fit parameters of  $\unit[12_{-4}^{+6}]{^\circ}$ and $\unit[7_{-3}^{+4}]{\%}$ for \textit{Swift}-BAT, $\unit[13_{-4}^{+7}]{^\circ}$ and $\unit[16_{-7}^{+8}]{\%}$ for 2MRS.

The different degrees of anisotropy obtained from each catalog can be understood from Fig.~\ref{Fig:Maps}, top, showing a UHECR hotspot in the direction of the Centaurus~A / M~83 / NGC~4945 group. The $\gamma$AGN model ($>\unit[60]{EeV}$) and \textit{Swift}-BAT model ($>\unit[39]{EeV}$) are dominated by Centaurus~A, which is $\unit[7]{^\circ}$ and $\unit[13]{^\circ}$ away from NGC~4945 and M~83, respectively. The starburst model additionally captures the UHECR excess close to the Galactic South pole, interpreted as contributions from NGC~1068 and NGC~253, yielding an increase in the anisotropy signal from ${\sim}3$ to $\unit[4]{\sigma}$. Additional diffuse contributions from clustered sources in the 2MRS catalog are not favored by the data, resulting in the smaller deviation from isotropy.

\begin{deluxetable*}{cccccccccc}
\tablecaption{Results - Scenario A\label{tab:Stats}}
\tablehead{
\colhead{Test} & \colhead{Null} & \colhead{Threshold} & \colhead{TS}  & \colhead{Local p-value} & \colhead{Post-trial} & \colhead{1-sided} &\colhead{AGN/other} &\colhead{SBG}&\colhead{Search} \\
\colhead{hypothesis} & \colhead{hypothesis} & \colhead{energy\tablenotemark{a} } & \colhead{}  & \colhead{$\mathcal{P}_{\chi^2}({\rm TS},2)$} & \colhead{p-value} & \colhead{significance} &\colhead{fraction} &\colhead{fraction}&\colhead{radius}}
\startdata
SBG  + ISO       & ISO & $\unit[39]{EeV}$  & 24.9 & $3.8\times10^{-6}$ & $3.6\times10^{-5}$  & 4.0\,$\sigma$ & N/A & $\unit[9.7]{\%}$ & $\unit[12.9]{^\circ}$ \\
$\gamma$AGN + SBG  + ISO & $\gamma$AGN + ISO  & $\unit[39]{EeV}$  & 14.7 & N/A            & $1.3\times10^{-4}$  &  3.7\,$\sigma$& $\unit[0.7]{\%} $ & $\unit[8.7]{\%}$ & $\unit[12.5]{^\circ}$ \\
&     &          &      &                    &                     &               & & & \\
$\gamma$AGN + ISO     & ISO &  $\unit[60]{EeV}$ & 15.2 & $5.1\times10^{-4}$ & $3.1\times10^{-3}$ & 2.7\,$\sigma$ & $\unit[6.7]{\%}$ & N/A & $\unit[6.9]{^\circ}$ \\
$\gamma$AGN + SBG  + ISO & SBG  + ISO  &  $\unit[60]{EeV}$ & 3.0  & N/A            & $0.08$              & 1.4\,$\sigma$ & $\unit[6.8]{\%} $ &  $\unit[0.0]{\%}$\tablenotemark{b} & $\unit[7.0]{^\circ}$ \\
\hline
{\it Swift}-BAT  + ISO       & ISO & $\unit[39]{EeV}$  & 18.2 & $1.1\times10^{-4}$ & $8.0\times10^{-4}$  & 3.2\,$\sigma$  & $\unit[6.9]{\%}$ & N/A & $\unit[12.3]{^\circ}$ \\
{\it Swift}-BAT + SBG  + ISO &  {\it Swift}-BAT + ISO  & $\unit[39]{EeV}$  & 7.8 & N/A & $5.1\times 10^{-3}$ & 2.6\,$\sigma$& $\unit[2.8]{\%} $ & $\unit[7.1]{\%}$ & $\unit[12.6]{^\circ}$ \\
&     &          &      &                    &                     &               & & & \\
2MRS  + ISO       & ISO & $\unit[38]{EeV}$  & 15.1 & $5.2\times10^{-4}$ & $3.3\times10^{-3}$  & 2.7\,$\sigma$ & $\unit[15.8]{\%}$ & N/A & $\unit[13.2]{^\circ}$ \\
2MRS + SBG  + ISO &  2MRS + ISO  & $\unit[39]{EeV}$  & 10.4 & N/A & $1.3\times 10^{-3}$  & 3.0\,$\sigma$& $\unit[1.1]{\%} $ & $\unit[8.9]{\%}$ & $\unit[12.6]{^\circ}$\\
\enddata
\tablenotetext{a}{For composite model studies, no scan over the threshold energy is performed.}
\tablenotetext{b}{Maximum TS reached at the boundary of the parameter space.}
\tablenotetext{}{ISO: isotropic model.}
\end{deluxetable*}

\subsection{Composite models against single populations}

To compare the two distinct gamma-ray populations  above their respective preferred thresholds, we investigate a composite model combining contributions from $\gamma$AGNs and SBGs, adopting a single search radius and leaving the fraction of events from each population free. The TS in this case is the difference between the maximum likelihood of the combined model and that of the null hypothesis of a single population at the selected energy threshold. The parameter added to the more complex model results in a $\chi^2$ distribution with one degree of freedom.

The best-fit anisotropic fractions obtained for the composite model (free search radius) are shown in Fig.~\ref{Fig:TSmaps}, right. Above $\unit[39]{EeV}$, the $\gamma$AGN-only model is disfavored by $\unit[3.7]{\sigma}$ relative to a combined model with a $\unit[9]{\%}$ contribution from SBGs and $\unit[1]{\%}$ contribution from $\gamma$AGN. Above $\unit[60]{EeV}$, the TS obtained with the composite model is not significantly higher than what is obtained by either model. This is illustrated in Fig.~\ref{Fig:TSmaps}, right, by the agreement at the $\unit[1]{\sigma}$ level of a model including $\unit[0]{\%}$ SBGs / $\unit[7]{\%}$ $\gamma$AGNs with a model including $\unit[13]{\%}$ SBGs / $\unit[0]{\%}$ $\gamma$AGNs above $\unit[60]{EeV}$.

As summarized in Table~\ref{tab:Stats}, composite models including SBGs and either 2MRS or \textit{Swift}-BAT sources best match the data above $\unit[39]{EeV}$ for $\unit[9-7]{\%}$ fractions of events associated to SBGs and $\unit[1-3]{\%}$ to the flux-limited samples. A $\unit[3.0-2.6]{\sigma}$ advantage is found for the composite models including SBGs with respect to the 2MRS-only and \textit{Swift}-BAT-only models.

\section{Discussion}

We have compared the arrival directions of UHECRs detected at the Pierre Auger Observatory with two distinct gamma-ray samples and two flux-limited samples of extragalactic sources. Our comparison with SBGs shows that isotropy of UHECRs is disfavored with $\unit[4.0]{\sigma}$ confidence, accounting for the two free parameters and including the penalty for scanning over energy thresholds.  This is the most significant evidence found in this study for anisotropy of UHECRs on an intermediate angular scale. It should be noted, however, that numerous anisotropy studies have been conducted with data from the  Observatory, not only those that have been published by the Collaboration.  There is no rigorous way to evaluate a statistical penalty for other searches. 

The pattern of UHECR arrival directions is best matched by a model in which about $\unit[10]{\%}$ of those cosmic rays arrive from directions that are clustered around the directions of bright, nearby SBGs.We evaluated the possibility of additional contributions from nearby $\gamma$AGNs, such as Centaurus~A, and from more distant sources through a comparison with samples tracing the distribution of extragalactic matter. We find that the contribution from SBGs to the indication of anisotropy is larger than that of the alternative catalogs tested. Nonetheless, caution is required in identifying SBGs as the preferred sources prior to understanding the impact of bulk magnetic deflections.

The sky maps used in this analysis are derived without incorporating any effects of the extragalactic or Galactic magnetic fields and winds \citep[e.g.][]{2011ApJ...738..192P,2012ApJ...761L..11J,2015ASTRP...2...39B}.  In particular, the arrival directions of UHECRs from a given source are modeled by a symmetric Fisher distribution centered on the source position. We checked the plausibility of the best-fit search radius obtained above $\unit[39]{EeV}$ by simulating sky maps passed through the Galactic magnetic field from \cite{2012ApJ...761L..11J}, including a random component with a coherence length of $\unit[60]{pc}$ as in \cite{2016APh....85...54E}. For large deflections, UHECRs from a given SBG can leak in the direction of a neighboring source. The three composition scenarios discussed in Sec.~\ref{Sec:compo} yield reconstructed search radii of $5-25^\circ$,  bracketing the observed radius of $13^\circ$. The agreement is considered satisfactory given the uncertainties in our knowledge of the composition above $\unit[39]{EeV}$ and of the deflections by the Galactic magnetic field \citep{2017arXiv170702339U}. Further studies aiming at possibly improving the model maps including deflections are underway.

\begin{figure*}
\includegraphics[width=0.48\textwidth]{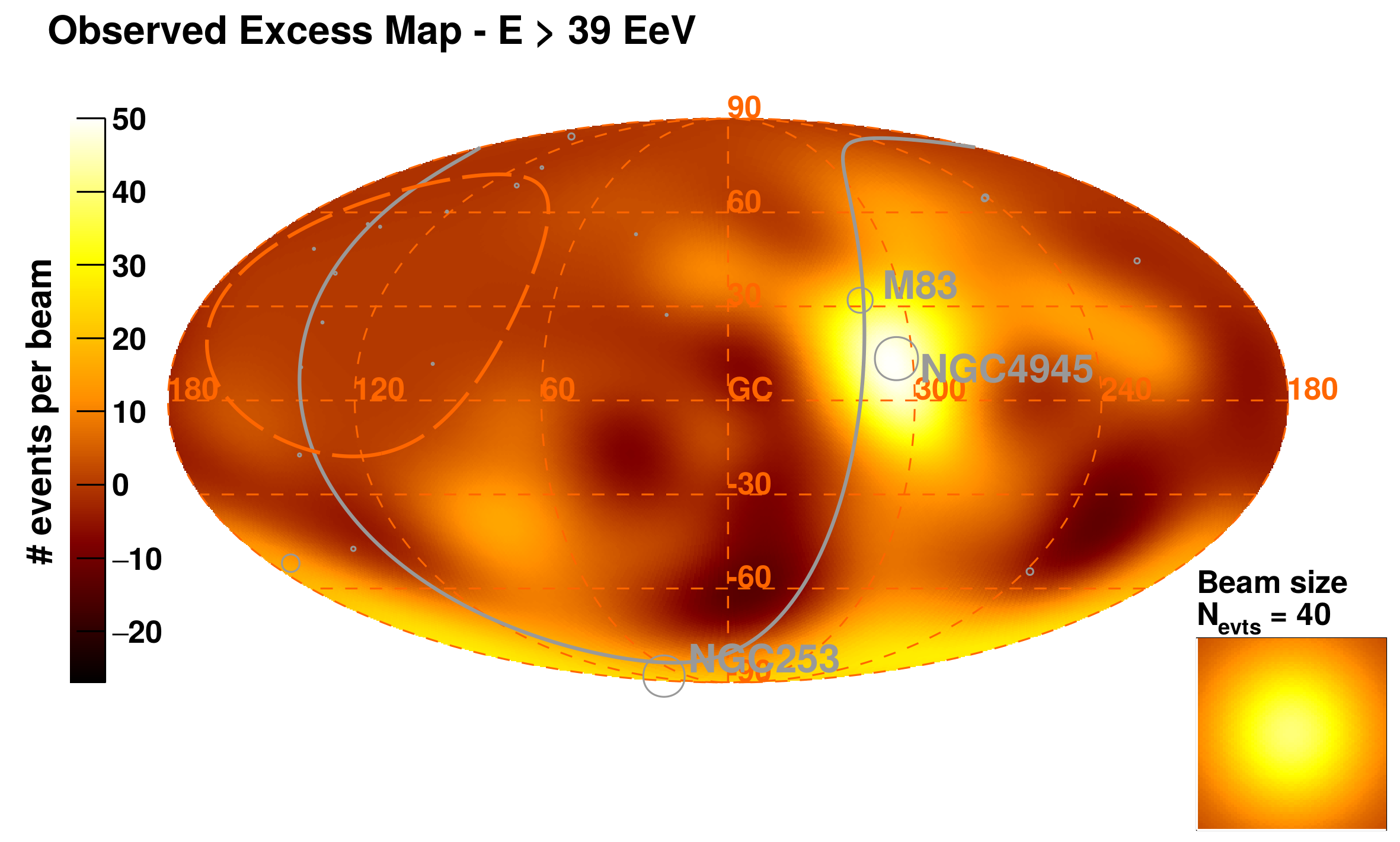}\hfill\includegraphics[width=0.48\textwidth]{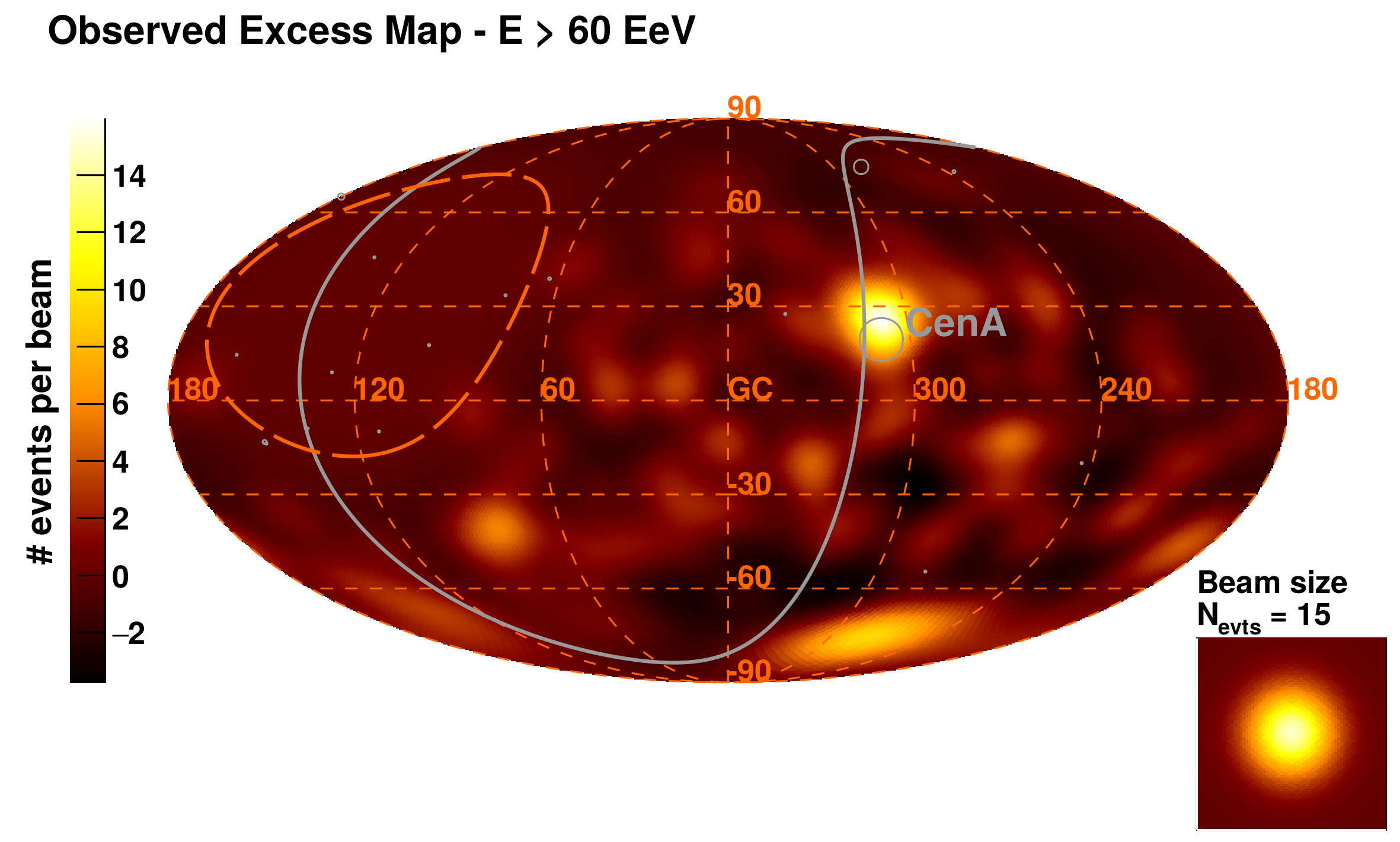}\\
\includegraphics[width=0.48\textwidth]{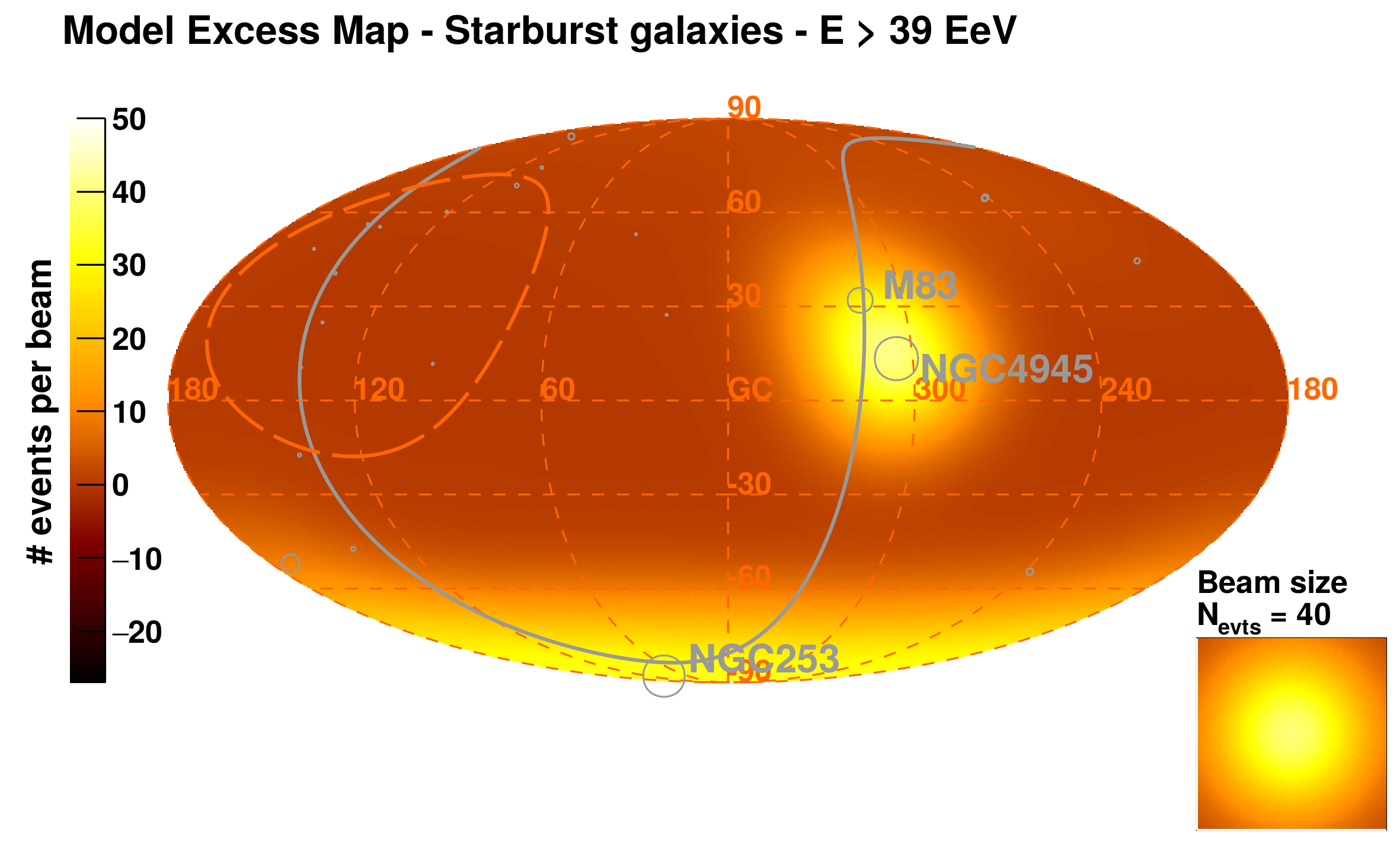}\hfill\includegraphics[width=0.48\textwidth]{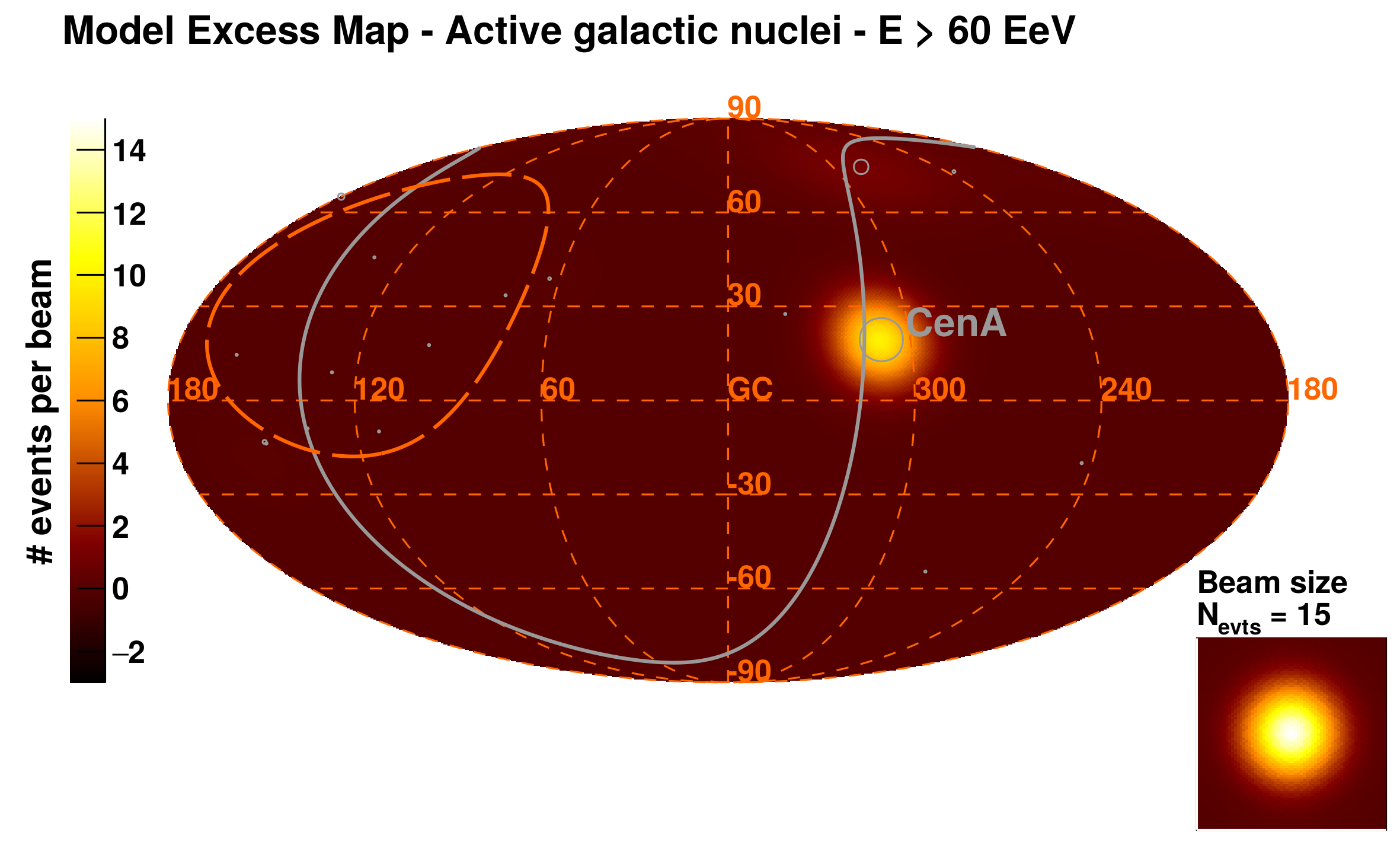}\\
\includegraphics[width=0.48\textwidth]{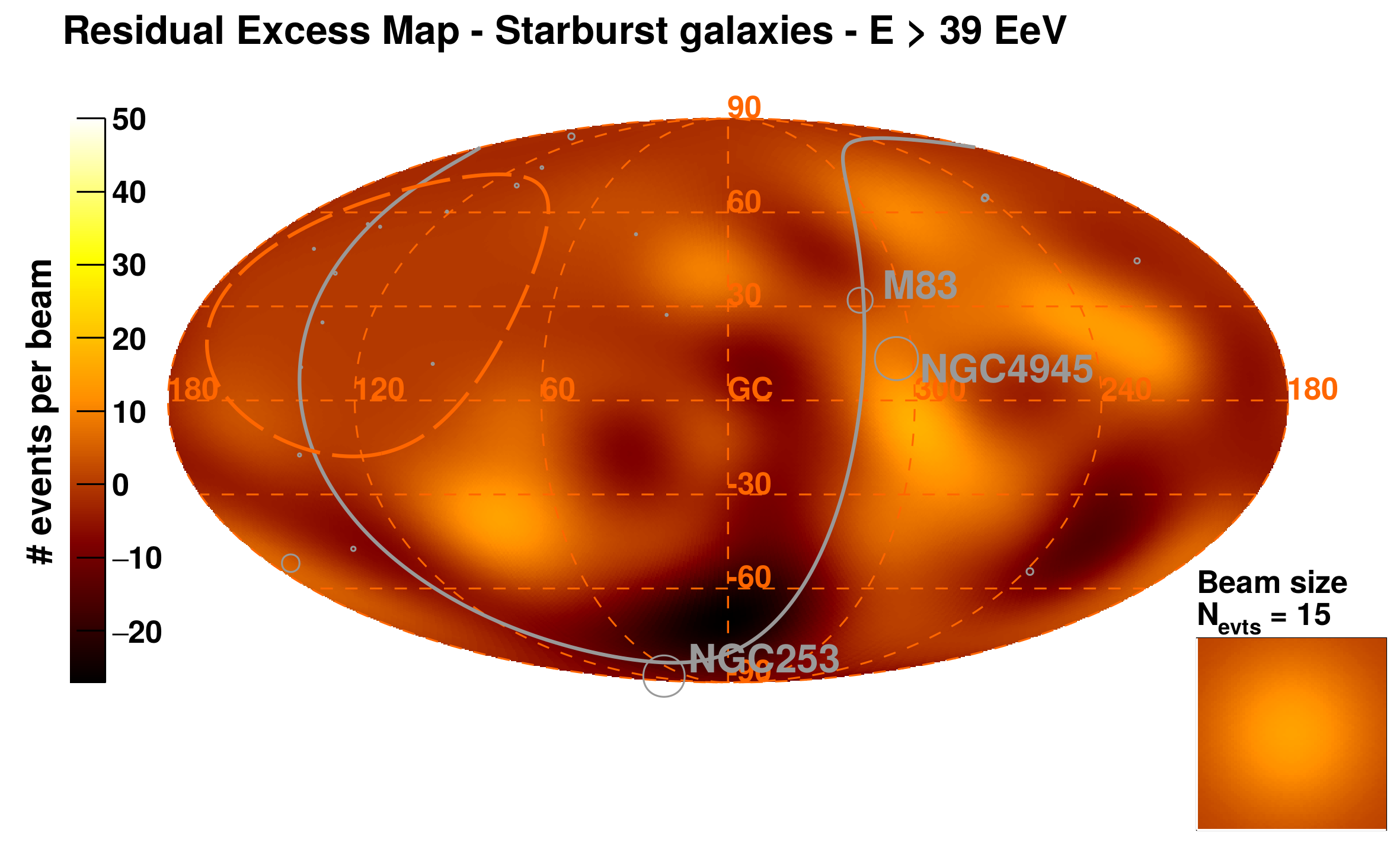}\hfill\includegraphics[width=0.48\textwidth]{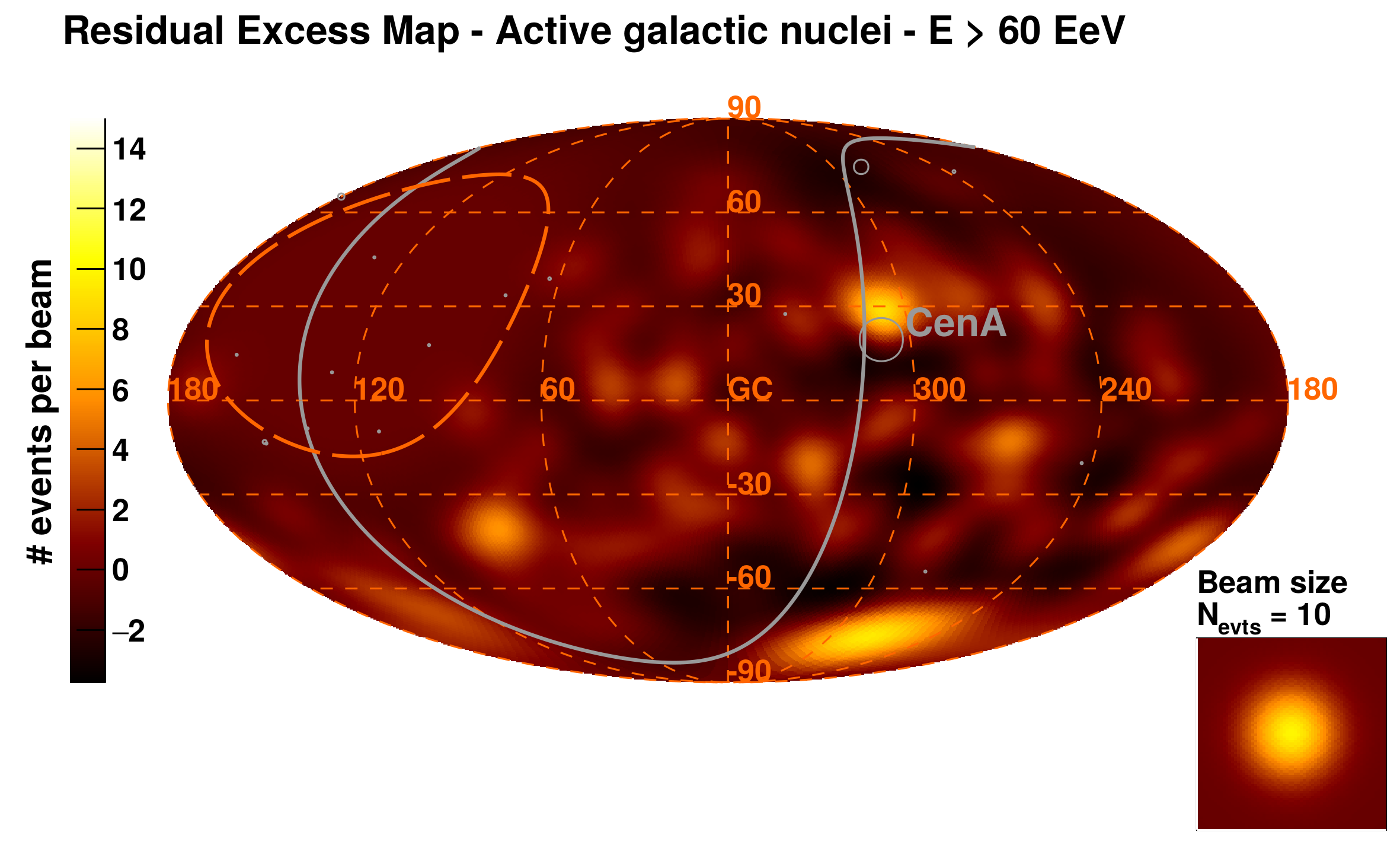}\\
\includegraphics[width=0.48\textwidth]{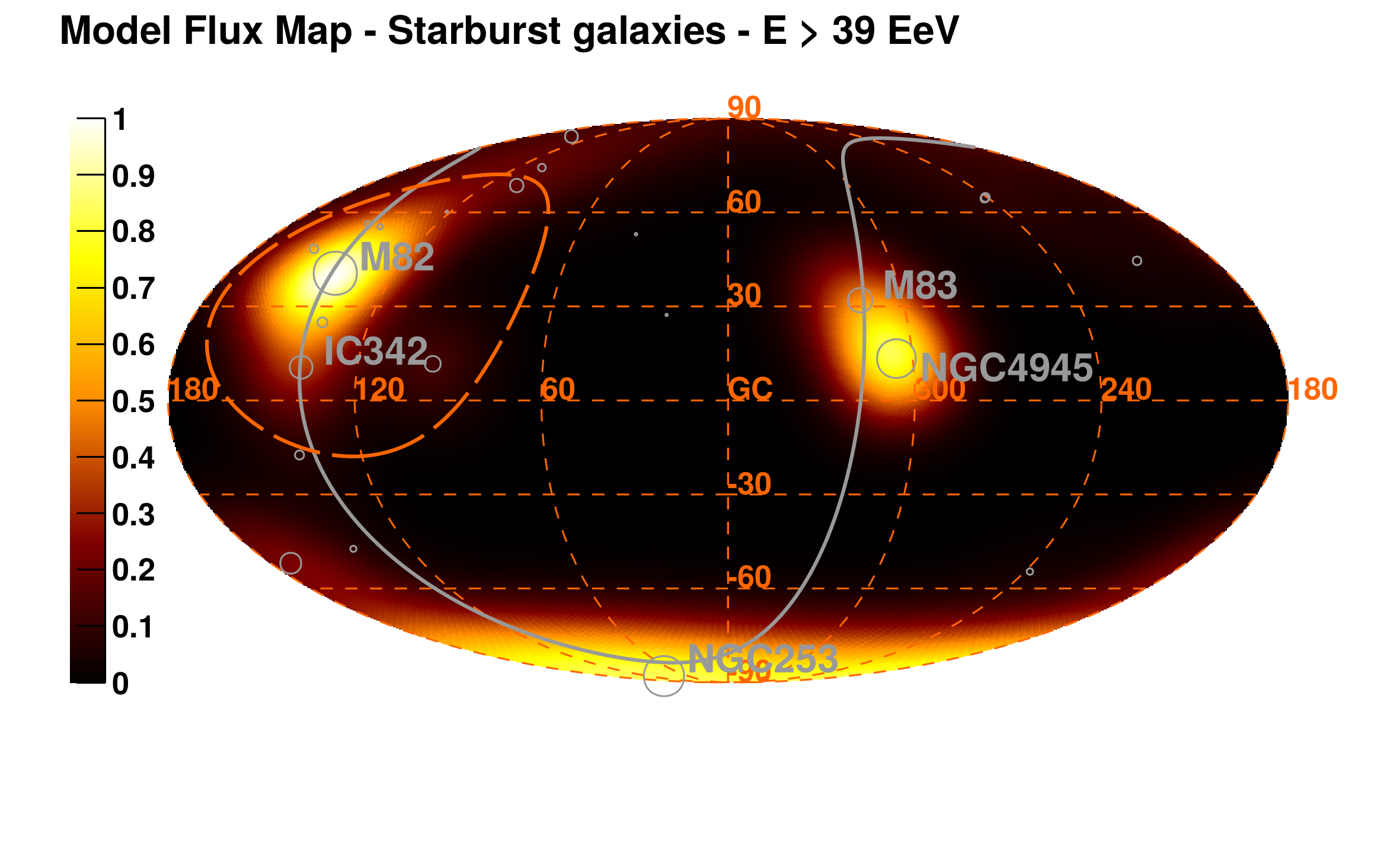}\hfill\includegraphics[width=0.48\textwidth]{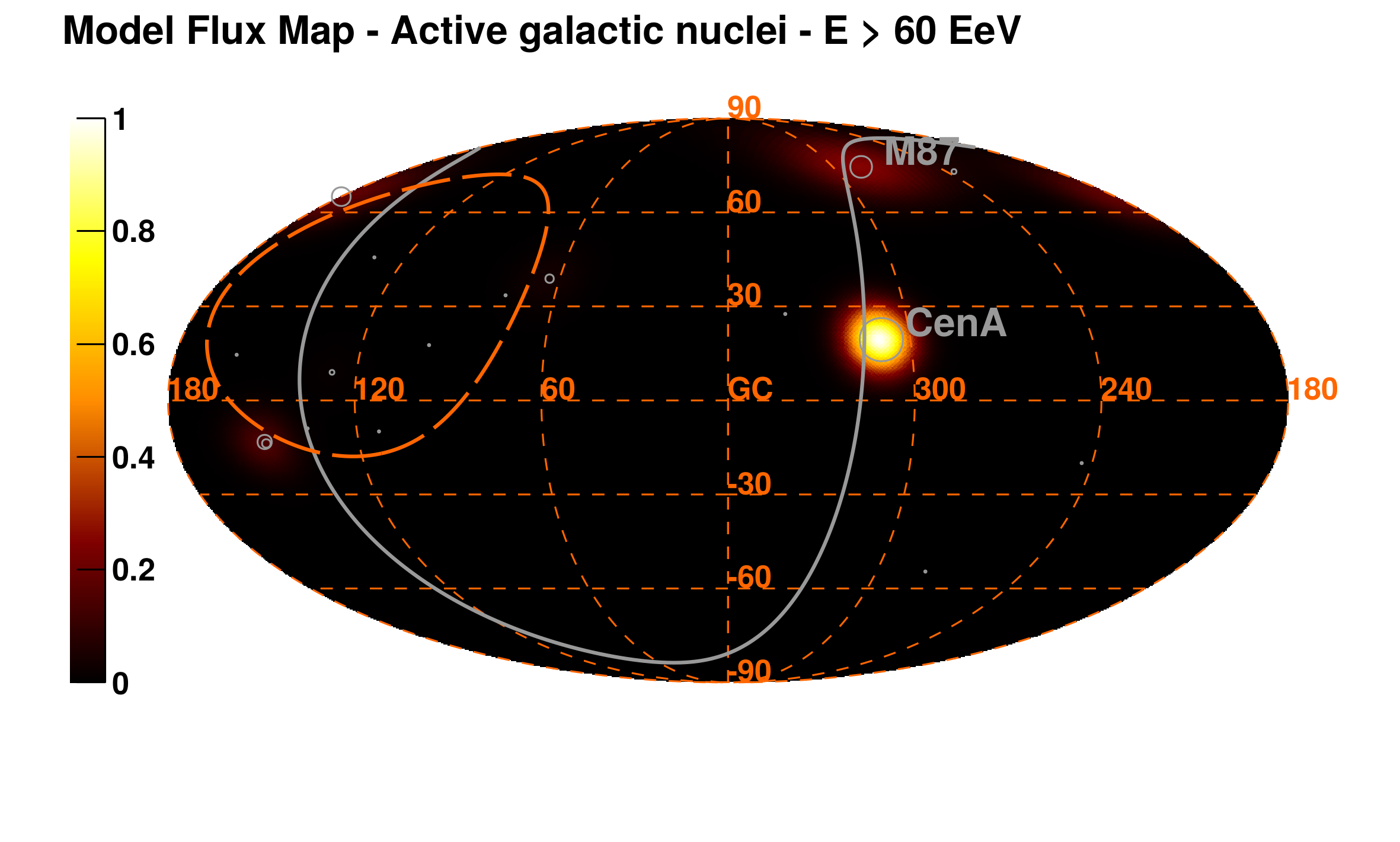}
\caption{\textit{Top} to \textit{Bottom}: Observed excess map - Model excess map - Residual map - Model flux map, for the best-fit parameters obtained with SBGs above $\unit[39]{EeV}$ (\textit{Left}) and $\gamma$AGNs above $\unit[60]{EeV}$ (\textit{Right}). The excess maps (best-fit isotropic component subtracted) and residual maps (observed minus model) are smeared at the best-fit angular scale. The color scale indicates the number of events per smearing beam (see inset). The model flux map corresponds to a uniform full-sky exposure. The supergalactic plane is shown as a solid gray line. An orange dashed line delimits the field of view of the array.}
\label{Fig:Maps}
\end{figure*}

It can be seen in Fig.~\ref{Fig:Maps}, bottom, that M~82 is expected to be one of the dominant sources in the full-sky starburst model presented here.  Its declination of ${\sim}\unit[70]{^\circ}$\,N is outside the exposure of the Observatory but is covered in the Northern Hemisphere by the Telescope Array~\citep{TA}. As noted e.g.\ by \cite{2014ApJ...794..126F} and \cite{2016PhRvD..93d3011H}, the excess of events observed at the Telescope Array~\citep{TAHotSpot} has some overlap with the position of M~82, as well as with the position of the blazar Mkn~421 that would be a bright Northern source in a low-attenuation scenario. 

An analysis of full-sky data from the Pierre Auger Observatory and the Telescope Array may provide a more powerful test of the starburst and AGN models by probing all production regions of UHECRs.  Combining the data is complicated, however, due to the spurious anisotropies that may be induced by possible mismatches in the relative exposures and/or systematic differences in the nominal energy scales used at each observatory. First attempts to conduct such surveys are being made~\citep{AugerTAUHECR16}.

Additional exposure will bring better constraints on the brightest sources. At the same time, an instrumentation upgrade of the Observatory is being deployed on the water-Cherenkov detectors adding a planar plastic scintillator of $\unit[4]{m^2}$ area to each of them~\citep{2016arXiv160403637T}. The upgrade will provide mass-sensitive observables for each shower enabling charge-discriminated studies with a duty cycle of nearly $\unit[100]{\%}$. Excluding highly charged nuclei from the analysis could eliminate a quasi-isotropic background that may mask the signature of individual sources imprinted by protons and other low-charge nuclei.  

Finally, a large-scale dipolar anisotropy has been discovered above $\unit[8]{EeV}$ \citep{dipole}. While a direct connection between the large and intermediate angular-scale patterns has not yet been identified, the emergence of anisotropies at ultra-high energy will certainly trigger further investigations of scenarios underlying the production of UHECRs.

\acknowledgments
The successful installation, commissioning, and operation of the Pierre Auger Observatory would not have been possible without the strong commitment from the technical and administrative staff in Malarg\"ue, and the financial support from a number of funding agencies in the participating countries, listed at \url{https://www.auger.org/index.php/about-us/funding-agencies}.

\bibliographystyle{yahapj}
\bibliography{ms}

\ 
\newpage
\begin{deluxetable*}{ccccccc}
\tablecaption{Populations investigated \label{tab:Pop}}
\tabletypesize{\footnotesize}
\tablehead{
\colhead{SBGs} & \colhead{l [$^\circ$]}            & \colhead{b [$^\circ$]}             & \colhead{Distance\tablenotemark{a} [Mpc]} & \colhead{Flux weight [\%]}        & \colhead{Attenuated weight: A / B / C [\%]}      & \colhead{\% contribution\tablenotemark{b}: A / B / C [\%]}} 
\startdata
NGC 253  	&	97.4	&	-88	&	2.7	&	13.6	&	20.7	/	18.0	/	16.6	&	35.9	/	32.2	/	30.2	\\
M82  	&	141.4	&	40.6	&	3.6	&	18.6	&	24.0	/	22.3	/	21.4	&	0.2	/	0.1	/	0.1	\\
NGC 4945  	&	305.3	&	13.3	&	4	&	16	&	19.2	/	18.3	/	17.9	&	39.0	/	38.4	/	38.3	\\
M83  	&	314.6	&	32	&	4	&	6.3	&	7.6	/	7.2	/	7.1	&	13.1	/	12.9	/	12.9	\\
IC 342  	&	138.2	&	10.6	&	4	&	5.5	&	6.6	/	6.3	/	6.1	&	0.1	/	0.0	/	0.0	\\
NGC 6946  	&	95.7	&	11.7	&	5.9	&	3.4	&	3.2	/	3.3	/	3.5	&	0.1	/	0.1	/	0.1	\\
NGC 2903  	&	208.7	&	44.5	&	6.6	&	1.1	&	0.9	/	1.0	/	1.1	&	0.6	/	0.7	/	0.7	\\
NGC 5055  	&	106	&	74.3	&	7.8	&	0.9	&	0.7	/	0.8	/	0.9	&	0.2	/	0.2	/	0.2	\\
NGC 3628  	&	240.9	&	64.8	&	8.1	&	1.3	&	1.0	/	1.1	/	1.2	&	0.8	/	0.9	/	1.1	\\
NGC 3627  	&	242	&	64.4	&	8.1	&	1.1	&	0.8	/	0.9	/	1.1	&	0.7	/	0.8	/	0.9	\\
NGC 4631  	&	142.8	&	84.2	&	8.7	&	2.9	&	2.1	/	2.4	/	2.7	&	0.8	/	0.9	/	1.1	\\
M51  	&	104.9	&	68.6	&	10.3	&	3.6	&	2.3	/	2.8	/	3.3	&	0.3	/	0.4	/	0.5	\\
NGC 891  	&	140.4	&	-17.4	&	11	&	1.7	&	1.1	/	1.3	/	1.5	&	0.2	/	0.3	/	0.3	\\
NGC 3556  	&	148.3	&	56.3	&	11.4	&	0.7	&	0.4	/	0.6	/	0.6	&	0.0	/	0.0	/	0.0	\\
NGC 660  	&	141.6	&	-47.4	&	15	&	0.9	&	0.5	/	0.6	/	0.8	&	0.4	/	0.5	/	0.6	\\
NGC 2146  	&	135.7	&	24.9	&	16.3	&	2.6	&	1.3	/	1.7	/	2.0	&	0.0	/	0.0	/	0.0	\\
NGC 3079  	&	157.8	&	48.4	&	17.4	&	2.1	&	1.0	/	1.4	/	1.5	&	0.1	/	0.1	/	0.1	\\
NGC 1068 	&	172.1	&	-51.9	&	17.9	&	12.1	&	5.6	/	7.9	/	9.0	&	6.4	/	9.4	/	10.9	\\
NGC 1365  	&	238	&	-54.6	&	22.3	&	1.3	&	0.5	/	0.8	/	0.8	&	0.9	/	1.5	/	1.6	\\
Arp 299  	&	141.9	&	55.4	&	46	&	1.6	&	0.4	/	0.7	/	0.6	&	0.0	/	0.0	/	0.0	\\
Arp 220  	&	36.6	&	53	&	80	&	0.8	&	0.1	/	0.3	/	0.2	&	0.0	/	0.2	/	0.1	\\
NGC 6240  	&	20.7	&	27.3	&	105	&	1	&	0.1	/	0.3	/	0.1	&	0.1	/	0.3	/	0.1	\\
Mkn 231  	&	121.6	&	60.2	&	183	&	0.8	&	0.0	/	0.1	/	0.0	&	0.0	/	0.0	/	0.0	\\
\cutinhead{$\gamma$AGNs}
  Cen A Core 	&	309.6	&	19.4	&	3.7	&	0.8	&	60.5	/	14.6	/	40.4	&	86.8	/	56.3	/	71.5	\\
M 87 	&	283.7	&	74.5	&	18.5	&	1	&	15.3	/	7.1	/	29.5	&	9.7	/	12.1	/	23.1	\\
NGC 1275 	&	150.6	&	-13.3	&	76	&	2.2	&	6.6	/	6.1	/	7.5	&	0.7	/	1.6	/	1.0	\\
IC 310 	&	150.2	&	-13.7	&	83	&	1	&	2.3	/	2.4	/	2.6	&	0.3	/	0.6	/	0.3	\\
3C 264 	&	235.8	&	73	&	95	&	0.5	&	0.8	/	1.0	/	0.8	&	0.4	/	1.3	/	0.5	\\
 TXS 0149+710 	&	127.9	&	9	&	96	&	0.5	&	0.7	/	0.9	/	0.7	&	0.0	/	0.0	/	0.0	\\
 Mkn 421 	&	179.8	&	65	&	136	&	54	&	11.4	/	48.3	/	14.7	&	1.8	/	19.1	/	2.8	\\
 PKS 0229-581 	&	280.2	&	-54.6	&	140	&	0.5	&	0.1	/	0.5	/	0.1	&	0.2	/	2.0	/	0.3	\\
 Mkn 501 	&	63.6	&	38.9	&	148	&	20.8	&	2.3	/	15.0	/	3.6	&	0.3	/	5.2	/	0.6	\\
 1ES 2344+514 	&	112.9	&	-9.9	&	195	&	3.3	&	0.0	/	1.0	/	0.1	&	0.0	/	0.0	/	0.0	\\
 Mkn 180 	&	131.9	&	45.6	&	199	&	1.9	&	0.0	/	0.5	/	0.0	&	0.0	/	0.0	/	0.0	\\
 1ES 1959+650 	&	98	&	17.7	&	209	&	6.8	&	0.0	/	1.7	/	0.1	&	0.0	/	0.0	/	0.0	\\
 AP Librae 	&	340.7	&	27.6	&	213	&	1.7	&	0.0	/	0.4	/	0.0	&	0.0	/	1.3	/	0.0	\\
 TXS 0210+515 	&	135.8	&	-9	&	218	&	0.9	&	0.0	/	0.2	/	0.0	&	0.0	/	0.0	/	0.0	\\
 GB6 J0601+5315 	&	160	&	14.6	&	232	&	0.4	&	0.0	/	0.1	/	0.0	&	0.0	/	0.0	/	0.0	\\
 PKS 0625-35 	&	243.4	&	-20	&	245	&	1.3	&	0.0	/	0.1	/	0.0	&	0.0	/	0.5	/	0.0	\\
 I Zw 187 	&	77.1	&	33.5	&	247	&	2.3	&	0.0	/	0.2	/	0.0	&	0.0	/	0.0	/	0.0	\\
\enddata
\tablenotetext{a}{A standard, flat $\Lambda$CDM model ($h_0=0.7$, $\Omega_{\rm M}=0.3$) is assumed. Distances of SBGs are based on \cite{2012ApJ...755..164A}, accounting for a small difference in $h_0$. Distances of $\gamma$AGNs are based on their redshifts, except for the nearby Cen~A \citep{2013AJ....146...86T}.}
\tablenotetext{b}{\% contributions account for the directional exposure of the array.}
\end{deluxetable*}

\ 
\newpage

\onecolumngrid
\centering
\section*{Authors}

\textsc{A.~Aab\textsuperscript{77}},
\textsc{P.~Abreu\textsuperscript{69}},
\textsc{M.~Aglietta\textsuperscript{50,49}},
\textsc{I.F.M.~Albuquerque\textsuperscript{18}},
\textsc{I.~Allekotte\textsuperscript{1}},
\textsc{A.~Almela\textsuperscript{8,11}},
\textsc{J.~Alvarez Castillo\textsuperscript{65}},
\textsc{J.~Alvarez-Mu\~niz\textsuperscript{76}},
\textsc{G.A.~Anastasi\textsuperscript{41,43}},
\textsc{L.~Anchordoqui\textsuperscript{83}}
% ...

(\textsc{The Pierre Auger Collaboration})
\vspace{0.0em}

\textsuperscript{1}{\footnotesize{Centro At\'omico Bariloche and Instituto Balseiro (CNEA-UNCuyo-CONICET), San Carlos de Bariloche, Argentina}}\\
\textsuperscript{2}{\footnotesize{Centro de Investigaciones en L\'aseres y Aplicaciones, CITEDEF and CONICET, Villa Martelli, Argentina}}\\
\textsuperscript{3}{\footnotesize{Departamento de F\'\i{}sica and Departamento de Ciencias de la Atm\'osfera y los Oc\'eanos, FCEyN, Universidad de Buenos Aires and CONICET, Buenos Aires, Argentina}}\\
\textsuperscript{4}{\footnotesize{IFLP, Universidad Nacional de La Plata and CONICET, La Plata, Argentina}}\\
\textsuperscript{5}{\footnotesize{Instituto de Astronom\'\i{}a y F\'\i{}sica del Espacio (IAFE, CONICET-UBA), Buenos Aires, Argentina}}\\
\textsuperscript{6}{\footnotesize{Instituto de F\'\i{}sica de Rosario (IFIR) -- CONICET/U.N.R.\ and Facultad de Ciencias Bioqu\'\i{}micas y Farmac\'euticas U.N.R., Rosario, Argentina}}\\
\textsuperscript{7}{\footnotesize{Instituto de Tecnolog\'\i{}as en Detecci\'on y Astropart\'\i{}culas (CNEA, CONICET, UNSAM), and Universidad Tecnol\'ogica Nacional -- Facultad Regional Mendoza (CONICET/CNEA), Mendoza, Argentina}}\\
\textsuperscript{8}{\footnotesize{Instituto de Tecnolog\'\i{}as en Detecci\'on y Astropart\'\i{}culas (CNEA, CONICET, UNSAM), Buenos Aires, Argentina}}\\
\textsuperscript{9}{\footnotesize{Observatorio Pierre Auger, Malarg\"ue, Argentina}}\\
\textsuperscript{10}{\footnotesize{Observatorio Pierre Auger and Comisi\'on Nacional de Energ\'\i{}a At\'omica, Malarg\"ue, Argentina}}\\
\textsuperscript{11}{\footnotesize{Universidad Tecnol\'ogica Nacional -- Facultad Regional Buenos Aires, Buenos Aires, Argentina}}\\
\textsuperscript{12}{\footnotesize{University of Adelaide, Adelaide, S.A., Australia}}\\
\textsuperscript{13}{\footnotesize{Universit\'e Libre de Bruxelles (ULB), Brussels, Belgium}}\\
\textsuperscript{14}{\footnotesize{Centro Brasileiro de Pesquisas Fisicas, Rio de Janeiro, RJ, Brazil}}\\
\textsuperscript{15}{\footnotesize{Centro Federal de Educa\c{c}\~ao Tecnol\'ogica Celso Suckow da Fonseca, Nova Friburgo, Brazil}}\\
\textsuperscript{16}{\footnotesize{Universidade de S\~ao Paulo, Escola de Engenharia de Lorena, Lorena, SP, Brazil}}\\
\textsuperscript{17}{\footnotesize{Universidade de S\~ao Paulo, Instituto de F\'\i{}sica de S\~ao Carlos, S\~ao Carlos, SP, Brazil}}\\
\textsuperscript{18}{\footnotesize{Universidade de S\~ao Paulo, Instituto de F\'\i{}sica, S\~ao Paulo, SP, Brazil}}\\
\textsuperscript{19}{\footnotesize{Universidade Estadual de Campinas, IFGW, Campinas, SP, Brazil}}\\
\textsuperscript{20}{\footnotesize{Universidade Estadual de Feira de Santana, Feira de Santana, Brazil}}\\
\textsuperscript{21}{\footnotesize{Universidade Federal de Pelotas, Pelotas, RS, Brazil}}\\
\textsuperscript{22}{\footnotesize{Universidade Federal do ABC, Santo Andr\'e, SP, Brazil}}\\
\textsuperscript{23}{\footnotesize{Universidade Federal do Paran\'a, Setor Palotina, Palotina, Brazil}}\\
\textsuperscript{24}{\footnotesize{Universidade Federal do Rio de Janeiro, Instituto de F\'\i{}sica, Rio de Janeiro, RJ, Brazil}}\\
\textsuperscript{25}{\footnotesize{Universidade Federal do Rio de Janeiro (UFRJ), Observat\'orio do Valongo, Rio de Janeiro, RJ, Brazil}}\\
\textsuperscript{26}{\footnotesize{Universidade Federal Fluminense, EEIMVR, Volta Redonda, RJ, Brazil}}\\
\textsuperscript{27}{\footnotesize{Universidad Industrial de Santander, Bucaramanga, Colombia}}\\
\textsuperscript{28}{\footnotesize{Charles University, Faculty of Mathematics and Physics, Institute of Particle and Nuclear Physics, Prague, Czech Republic}}\\
\textsuperscript{29}{\footnotesize{Institute of Physics of the Czech Academy of Sciences, Prague, Czech Republic}}\\
\textsuperscript{30}{\footnotesize{Palacky University, RCPTM, Olomouc, Czech Republic}}\\
\textsuperscript{31}{\footnotesize{Institut de Physique Nucl\'eaire d'Orsay (IPNO), Universit\'e Paris-Sud, Univ.\ Paris/Saclay, CNRS-IN2P3, Orsay, France}}\\
\textsuperscript{32}{\footnotesize{Laboratoire de Physique Nucl\'eaire et de Hautes Energies (LPNHE), Universit\'es Paris 6 et Paris 7, CNRS-IN2P3, Paris, France}}\\
\textsuperscript{33}{\footnotesize{Laboratoire de Physique Subatomique et de Cosmologie (LPSC), Universit\'e Grenoble-Alpes, CNRS/IN2P3, Grenoble, France}}\\
\textsuperscript{34}{\footnotesize{Bergische Universit\"at Wuppertal, Department of Physics, Wuppertal, Germany}}\\
\textsuperscript{35}{\footnotesize{Karlsruhe Institute of Technology, Institut f\"ur Experimentelle Kernphysik (IEKP), Karlsruhe, Germany}}\\
\textsuperscript{36}{\footnotesize{Karlsruhe Institute of Technology, Institut f\"ur Kernphysik, Karlsruhe, Germany}}\\
\textsuperscript{37}{\footnotesize{Karlsruhe Institute of Technology, Institut f\"ur Prozessdatenverarbeitung und Elektronik, Karlsruhe, Germany}}\\
\textsuperscript{38}{\footnotesize{RWTH Aachen University, III.\ Physikalisches Institut A, Aachen, Germany}}\\
\textsuperscript{39}{\footnotesize{Universit\"at Hamburg, II.\ Institut f\"ur Theoretische Physik, Hamburg, Germany}}\\
\textsuperscript{40}{\footnotesize{Universit\"at Siegen, Fachbereich 7 Physik -- Experimentelle Teilchenphysik, Siegen, Germany}}\\
\textsuperscript{41}{\footnotesize{Gran Sasso Science Institute (INFN), L'Aquila, Italy}}\\
\textsuperscript{42}{\footnotesize{INAF -- Istituto di Astrofisica Spaziale e Fisica Cosmica di Palermo, Palermo, Italy}}\\
\textsuperscript{43}{\footnotesize{INFN Laboratori Nazionali del Gran Sasso, Assergi (L'Aquila), Italy}}\\
\textsuperscript{44}{\footnotesize{INFN, Sezione di Catania, Catania, Italy}}\\
\textsuperscript{45}{\footnotesize{INFN, Sezione di Lecce, Lecce, Italy}}\\
\textsuperscript{46}{\footnotesize{INFN, Sezione di Milano, Milano, Italy}}\\
\textsuperscript{47}{\footnotesize{INFN, Sezione di Napoli, Napoli, Italy}}\\
\textsuperscript{48}{\footnotesize{INFN, Sezione di Roma "Tor Vergata", Roma, Italy}}\\
\textsuperscript{49}{\footnotesize{INFN, Sezione di Torino, Torino, Italy}}\\
\textsuperscript{50}{\footnotesize{Osservatorio Astrofisico di Torino (INAF), Torino, Italy}}\\
\textsuperscript{51}{\footnotesize{Politecnico di Milano, Dipartimento di Scienze e Tecnologie Aerospaziali , Milano, Italy}}\\
\textsuperscript{52}{\footnotesize{Universit\`a del Salento, Dipartimento di Ingegneria, Lecce, Italy}}\\
\textsuperscript{53}{\footnotesize{Universit\`a del Salento, Dipartimento di Matematica e Fisica ``E.\ De Giorgi'', Lecce, Italy}}\\
\textsuperscript{54}{\footnotesize{Universit\`a dell'Aquila, Dipartimento di Scienze Fisiche e Chimiche, L'Aquila, Italy}}\\
\textsuperscript{55}{\footnotesize{Universit\`a di Catania, Dipartimento di Fisica e Astronomia, Catania, Italy}}\\
\textsuperscript{56}{\footnotesize{Universit\`a di Milano, Dipartimento di Fisica, Milano, Italy}}\\
\textsuperscript{57}{\footnotesize{Universit\`a di Napoli "Federico II", Dipartimento di Fisica ``Ettore Pancini``, Napoli, Italy}}\\
\textsuperscript{58}{\footnotesize{Universit\`a di Roma ``Tor Vergata'', Dipartimento di Fisica, Roma, Italy}}\\
\textsuperscript{59}{\footnotesize{Universit\`a Torino, Dipartimento di Fisica, Torino, Italy}}\\
\textsuperscript{60}{\footnotesize{Benem\'erita Universidad Aut\'onoma de Puebla, Puebla, M\'exico}}\\
\textsuperscript{61}{\footnotesize{Centro de Investigaci\'on y de Estudios Avanzados del IPN (CINVESTAV), M\'exico, D.F., M\'exico}}\\
\textsuperscript{62}{\footnotesize{Unidad Profesional Interdisciplinaria en Ingenier\'\i{}a y Tecnolog\'\i{}as Avanzadas del Instituto Polit\'ecnico Nacional (UPIITA-IPN), M\'exico, D.F., M\'exico}}\\
\textsuperscript{63}{\footnotesize{Universidad Aut\'onoma de Chiapas, Tuxtla Guti\'errez, Chiapas, M\'exico}}\\
\textsuperscript{64}{\footnotesize{Universidad Michoacana de San Nicol\'as de Hidalgo, Morelia, Michoac\'an, M\'exico}}\\
\textsuperscript{65}{\footnotesize{Universidad Nacional Aut\'onoma de M\'exico, M\'exico, D.F., M\'exico}}\\
\textsuperscript{66}{\footnotesize{Institute of Nuclear Physics PAN, Krakow, Poland}}\\
\textsuperscript{67}{\footnotesize{University of \L{}\'od\'z, Faculty of Astrophysics, \L{}\'od\'z, Poland}}\\
\textsuperscript{68}{\footnotesize{University of \L{}\'od\'z, Faculty of High-Energy Astrophysics,\L{}\'od\'z, Poland}}\\
\textsuperscript{69}{\footnotesize{Laborat\'orio de Instrumenta\c{c}\~ao e F\'\i{}sica Experimental de Part\'\i{}culas -- LIP and Instituto Superior T\'ecnico -- IST, Universidade de Lisboa -- UL, Lisboa, Portugal}}\\
\textsuperscript{70}{\footnotesize{``Horia Hulubei'' National Institute for Physics and Nuclear Engineering, Bucharest-Magurele, Romania}}\\
\textsuperscript{71}{\footnotesize{Institute of Space Science, Bucharest-Magurele, Romania}}\\
\textsuperscript{72}{\footnotesize{University Politehnica of Bucharest, Bucharest, Romania}}\\
\textsuperscript{73}{\footnotesize{Center for Astrophysics and Cosmology (CAC), University of Nova Gorica, Nova Gorica, Slovenia}}\\
\textsuperscript{74}{\footnotesize{Experimental Particle Physics Department, J.\ Stefan Institute, Ljubljana, Slovenia}}\\
\textsuperscript{75}{\footnotesize{Universidad de Granada and C.A.F.P.E., Granada, Spain}}\\
\textsuperscript{76}{\footnotesize{Universidad de Santiago de Compostela, Santiago de Compostela, Spain}}\\
\textsuperscript{77}{\footnotesize{IMAPP, Radboud University Nijmegen, Nijmegen, The Netherlands}}\\
\textsuperscript{78}{\footnotesize{KVI -- Center for Advanced Radiation Technology, University of Groningen, Groningen, The Netherlands}}\\
\textsuperscript{79}{\footnotesize{Nationaal Instituut voor Kernfysica en Hoge Energie Fysica (NIKHEF), Science Park, Amsterdam, The Netherlands}}\\
\textsuperscript{80}{\footnotesize{Stichting Astronomisch Onderzoek in Nederland (ASTRON), Dwingeloo, The Netherlands}}\\
\textsuperscript{81}{\footnotesize{Case Western Reserve University, Cleveland, OH, USA}}\\
\textsuperscript{82}{\footnotesize{Colorado School of Mines, Golden, CO, USA}}\\
\textsuperscript{83}{\footnotesize{Department of Physics and Astronomy, Lehman College, City University of New York, Bronx, NY, USA}}\\
\textsuperscript{84}{\footnotesize{Louisiana State University, Baton Rouge, LA, USA}}\\
\textsuperscript{85}{\footnotesize{Michigan Technological University, Houghton, MI, USA}}\\
\textsuperscript{86}{\footnotesize{New York University, New York, NY, USA}}\\
\textsuperscript{87}{\footnotesize{Northeastern University, Boston, MA, USA}}\\
\textsuperscript{88}{\footnotesize{Ohio State University, Columbus, OH, USA}}\\
\textsuperscript{89}{\footnotesize{Pennsylvania State University, University Park, PA, USA}}\\
\textsuperscript{90}{\footnotesize{University of Chicago, Enrico Fermi Institute, Chicago, IL, USA}}\\
\textsuperscript{91}{\footnotesize{University of Nebraska, Lincoln, NE, USA}}\\
\vspace{-0.5em}\textsuperscript{}{-----}\\\vspace{-0.5em}
\textsuperscript{1000}{\footnotesize{School of Physics and Astronomy, University of Leeds, Leeds, United Kingdom}}\\
\textsuperscript{1001}{\footnotesize{Max-Planck-Institut f\"ur Radioastronomie, Bonn, Germany}}\\
\textsuperscript{1002}{\footnotesize{also at Vrije Universiteit Brussels, Brussels, Belgium}}\\
\textsuperscript{1003}{\footnotesize{SUBATECH, \'Ecole des Mines de Nantes, CNRS-IN2P3, Universit\'e de Nantes, France}}\\
\textsuperscript{1004}{\footnotesize{Fermi National Accelerator Laboratory, USA}}\\
\textsuperscript{1005}{\footnotesize{Colorado State University, Fort Collins, CO}}\\
{\footnotesize{$\dagger$ Deceased}

\end{document}